\documentclass{article}
\title{Doubly Special Relativity in Position Space Starting from the 
Conformal Group.}
\author{A.A. Deriglazov\footnote{alexei@ice.ufjf.br ~ On leave of
absence from Dept. Math. Phys., Tomsk Polytechnical University,
Tomsk, Russia.}}
\date{Dept. de Matematica, ICE, Universidade Federal de Juiz de Fora,\\
MG, Brasil.}
\begin{document}
\maketitle
\large
\begin{abstract}
We propose version of doubly special relativity theory starting from  
position space. The version is based on deformation of ordinary 
Lorentz transformations due to the special conformal transformation. 
There is unique deformation  
which does not modify rotations. In contrast to the Fock-Lorentz 
realization (as well as to recent position-space proposals),  
maximum signal velocity is position (and observer) independent scale in 
our formulation by construction. The formulation admits one more invariant scale  
identified with radius of three-dimensional space-like hypersection 
of space-time. 
We present and discuss the Lagrangian action for geodesic motion of a particle 
on the DSR space. For the present formulation, one needs to distinguish 
the canonical (conjugated to $x^\mu$) momentum $p^\mu$ from the conserved 
energy-momentum.  
Deformed Lorentz transformations for $x^\mu$ induce complicated transformation law  
in space of canonical momentum. $p^\mu$ is not a conserved quantity and 
obeys to deformed dispersion relation. 
The conserved energy-momentum $P^\mu$ turns out to be different from the canonical 
one, in particular, $P^\mu$-space is equipped with nontrivial commutator. 
The nonlinear transformations for $x^\mu$  
induce the standard Lorentz transformations in 
space of $P^\mu$. It means, in particular, that composite rule 
for $P^\mu$ is ordinary sum. There is no problem of total momentum 
in the theory. $P^\mu$ obeys the 
standard energy-momentum relation (while has nonstandard dependence on velocity).

\end{abstract}

%{\bf PAC codes:} 0460D, 1130C, 1125 \\
%{\bf Keywords:} Doubly Special Relativity, 
%Deformed Energy-Momentum Relations, Lorentz Violating \\

\noindent
\newpage
\section{Introduction.}

Doubly special relativity (DSR) proposals [1-5] might be specified as the 
theories with underlying symmetry group being the Lorentz group
\footnote{For the early proposals, based on $\kappa$-Poincare algebra, see 
[7].}, but with 
kinematical predictions different from that of special relativity. It can be 
achieved by taking of some deformation of the Lorentz group realization     
in space of conserved energy-momentum. In particular, Magueijo-Smolin (MS)  
suggestion [2, 3] is to take the momentum space realization of the group in the form  
$\Lambda_U=U^{-1}\Lambda U$, where $\Lambda$ represents 
ordinary Lorentz transformation and $U(P^\mu)$ is some operator. 
Ordinary energy-momentum relation $(P^\mu)^2=-m^2$ is not invariant under 
the realization and is replaced by $[U(P^\mu)]^2=-m^2$. It suggests kinematical 
predictions different from that of special relativity. 
There is a number of 
attractive motivations for such a modification (see discussion in [1-5]), 
in particular, one can believe on DSR as an intermediate theory where the 
quantum gravity effects are presented even in the regime of neglible 
gravitational field [9, 3]. In turn, it implies that the formulation includes 
dimensional parameter ($U=U(P^\mu, \lambda))$ in such a way that one recovers  
special relativity 
in some limit. The parameter (or parameters, see the recent work [6]) turns  
out to be one more (in addition to speed of light) observer independent 
scale present in the formulation. The scale was identified with the Planck 
energy in [2, 3]. The emergence of a new scale was taken as the 
guiding principle for construction of different DSR models in a number of papers.
Modifications with various dimensional scales has been proposed [1-3, 4, 5, 6, 13]. 
In particular, in the work [6] it was discussed an algebraic construction which 
implies three scales $c, ~ E_p, ~ R$, with $E_p$ identified with the Planck energy 
and $R$ being the cosmological constant.

To complete the picture, it is desirable to find underlying space-time interpretation   
for the DSR kinematics, that is to construct position space realization of the Lorentz 
group which generates one or another DSR kinematics.  
Then one could be able to formulate dynamical problems on DSR space in 
the standard 
framework, starting from the action functional, which suggests physical 
interpretation of the results obtained in momentum space formulation. 
Actually, in this case the 
spaces of velocities, of canonical (conjugated to the position) momentum, the 
energy-momentum space as well as their properties 
and map of one to another can be obtained by direct computations 
(the issue being rather delicate question in the formulation with  
energy-momentum space as the starting 
point [8, 9, 10]). One expects also that the central problem of the DSR kinematics 
(the problem of total momentum for multi-particle system) can be clarified in the 
position space formulation. 

To find a  
position version for the given DSR kinematics one needs to decide, in fact, 
what is the relation among the energy-momentum $P^\mu$ and the position 
variables $x^\mu$ 
(as well as the canonical momenta $p^\mu$), the approach undertaken 
in [5, 3, 8, 9, 11, 13]. Let us enumerate some of the results.

Assuming coincidence of $P^\mu$ and $p^\mu$ 
(equivalently, invariance of $P dx$) [5, 10], one obtains the energy-momentum 
dependent Lorentz transformations in position space. 

In the algebraic approach [11],  
position version is encoded in the Poisson brackets of an algebra which unifies 
the Poincare algebra and the phase space one. It implies 
noncommutativity\footnote{Appearance of noncommutative geometry in the  
DSR framework might be starting point to treat the problem of Lorentz 
(rotational) invariance in different noncommutative quantum mechanical 
models [12].} of position variables [6]. 

For the MS kinematics it is possible to take  
ordinary Lorentz realization on $x^\mu$ and then to deform standard relation 
among $x$ and 
$P$ in some particular way [13]. Then the MS invariant and the MS transformations     
are generated on the momentum space from $\dot x^2=-m^2c^2, 
~ x^{'\mu}=\Lambda^\mu{}_\nu x^\nu$, which gives a consistent picture in 
one-particle sector. Being quite simple, this point of view seems to be unreasonable, 
mainly due to the fact that it is difficult to construct an addition 
rule with acceptable physical properties in the multi-particle sector of the theory
(see [13] for detailed discussion). 

Besides the nonlinear MS transformations, the MS energy - momentum relation  
is invariant also under some inhomogeneous linear transformations [13]. 
The latter are induced starting from linearly realized Lorentz group in 
five-dimensional position space. Fer the case, there are different possibilities 
to relate new scale with fundamental constants. In particular, identification 
with vacuum energy suggests emergence of minimum quantum of mass [13]. 

The abovementioned works are devoted to search for space-time interpretation 
of a given DSR kinematics. Instead of this, one can ask on reasonable deformations 
of the 
Lorentz group realization in position space without reference on a particular 
DSR kinematics [14, 5]. We follow this line in the present work.  
We propose deformation of the Lorentz transformations  
based on the conformal group. By construction, maximum signal velocity is observer 
(and space-time) independent scale of the formulation, the latter is described 
in Section 2. In Section 3 we discuss geodesic motion of a particle,  
with the Lagrangian action being invariant interval of the DSR space. Kinematics 
corresponding to the theory is constructed and discussed in some details. 
In particular, the present formulation turns out to be free of the problem of total 
momentum in many-particle sector.    
  
\section{Deformation of the Lorentz transformations due to the special conformal 
transformation.}

In this Section we motivate that the conformal group in four dimensions 
seem to be an appropriate framework to formulate the DSR models in  
position space\footnote{It was observed in [18] that the MS operator $U$ and the 
special conformal transformation (on momentum space) with $b^\mu=(l_p, 0, 0, 0)$ 
coincide on the surface $p^2=0$.}.
  
In ordinary special relativity the requirement of invariance of the Minkowski 
interval: $ds^{'2}=ds^2$ immediately leads to the observer independent 
scale $|v^i|=c$. To construct a theory with one more scale, the invariance 
condition seems to be too restrictive. Actually, the most general transformations 
$x^\mu\longrightarrow x^{'\mu}(x^\nu)$ which preserve the interval are known to 
be the Lorentz transformations in the standard realization [15] 
$x^{'\mu}=\Lambda^\mu{}_\nu x^\nu$, the latter does not admit one more invariant 
scale. So, one needs to relax the invariance condition keeping, as before, the 
speed of light invariant. It would be the case if $ds^2=0$ will imply 
$ds^{'2}=0$, which guarantees appearance of the invariant scale $c$ (in 
the case of linear relation $x^0=ct$). 

Thus, supposing existence of one more observer independent scale $R$, one assumes 
deformation of the invariance condition: $ds^{'2}=A(x,R)ds^2$, where 
$A \stackrel{R\rightarrow \infty}{\longrightarrow}1$. By construction, the maximum 
velocity remains the invariant scale of the formulation. In the limit 
$R\rightarrow \infty$ one obtains the ordinary special relativity theory. 

Complete symmetry 
group for the case is the conformal group (see, for example [16]). Besides 
the Lorentz transformations it 
consist of the dilatations $x^{'\mu}=\rho x^\mu$ and the special conformal 
transformations with the parameter $b^\mu$
\begin{eqnarray}\label{1}
SC_b: x^\mu\longrightarrow\frac{1}{\Omega}(x^\mu+b^\mu x^2), \cr 
\Omega(x, b)\equiv 1+2b x+b^2x^2.\quad
\end{eqnarray}
Similarly to the previous DSR proposals [2, 3, 5], let us deform the Lorentz group 
realization in accordance with the rule $\Lambda_{def}=U^{-1}\Lambda U$. We take 
the special conformal transformation\footnote{invariance under the 
complete conformal group leads to the massless particle, which is not of our interest 
here.} with some fixed $b^\mu$ being the similarity operator $U$
\begin{eqnarray}\label{2}
\Lambda_b\equiv (SC_b)^{-1}\Lambda (SC_b), \qquad \qquad \cr 
\Lambda_b: x^\mu\longrightarrow\frac{1}{G}\left[(\Lambda x)^\mu+
\left[(1-\Lambda)b\right]^\mu x^2\right], \quad \cr
G(x, b, \Lambda)\equiv 1-2b(1-\Lambda)x+2b(1-\Lambda)bx^2.
\end{eqnarray}
The above mentioned proportionality factor for the case is $A=G^{-2}$. 
The parameters $b^\mu$ can be 
further specified by the requirement that space rotations 
$\Lambda^\mu{}_\nu=( \Lambda^0{}_0 =1, ~  
\Lambda^0{}_i= \Lambda^i{}_0 =0, ~ \Lambda^i{}_j\equiv R^i{}_j, ~ 
R^T=R^{-1})$ are not deformed 
by $b^\mu$. Then the only choice is $b^\mu=(\lambda, 0, 0, 0)$, 
which gives the final form of the deformed Lorentz group realization
\begin{eqnarray}\label{3}
\Lambda_\lambda: x^\mu\longrightarrow\frac{1}{G}\left[(\Lambda x)^\mu+
(\delta^\mu{}_0-\Lambda^\mu{}_0)\lambda x^2\right], \qquad \\
G(x, \lambda, \Lambda)\equiv 1+2\lambda(x^0-\Lambda^0{}_\mu x^\mu)
-2\lambda^2(1-\Lambda^0{}_0)x^2.
\end{eqnarray}
Our convention for the Minkowski metric is $\eta_{\mu\nu}=(-, +, +, +)$.
One confirms now emergence of one more observer independent scale:  
there is exist unique vector $x^\mu$ with zero component unaltered by 
the transformations (\ref{3}). Namely, from the condition 
$x^{'0}=x^0$ one has the only solution $x^\mu=(R\equiv -\frac{1}{\lambda}, 0, 0, 0)$ 
(the latter turns out to be the fixed vector). Thus all  
observers should agree to identify $R$ as the invariant scale. Let us point 
that the transformations (\ref{3}) are not equivalent to either the 
Fock-Lorentz realization [15], or to 
recent DSR proposals (the realizations lead to varying speed of light).

Invariant interval under the transformations (\ref{3}) can be find by 
inspection of transformation properties of the following quantities:
\begin{eqnarray}\label{4}
dx^{'\mu}=\frac{1}{G^2}[((\Lambda dx)^\mu+
2\lambda(\delta^\mu{}_{0}-\Lambda^\mu{}_{0})(x dx))G- \qquad \qquad \cr
2\lambda(dx^0-\Lambda^0{}_{\nu}dx^\nu-2\lambda(1-\Lambda^0{}_{0})(x dx))
((\Lambda x)^\mu+\lambda(\delta^\mu{}_{0}-\Lambda^\mu{}_{0})x^2)],
\end{eqnarray}
\begin{eqnarray}\label{41}
(dx^{'\mu})^2=\frac{(dx^{\mu})^2}{G^2}, \qquad 
\tilde\Omega^{'}=\frac{\tilde\Omega}{G}, \qquad
\end{eqnarray}
where 
\begin{eqnarray}\label{5}
\tilde\Omega\equiv\Omega(-\lambda)=1+2\lambda x^0-\lambda^2x^2.
\end{eqnarray}
Then the quantity
\begin{eqnarray}\label{6}
ds^2=\frac{\eta_{\mu\nu}dx^\mu dx^\nu}{(1+2\lambda x^0-\lambda^2x^2)^2}
\equiv g_{\mu\nu}(x)dx^\mu dx^\nu,
\end{eqnarray}
represents the invariant interval. On the domain where the metric is 
non degenerated, the corresponding four dimensional scalar curvature is zero, 
while three-dimensional space-like slice $x^0=0$ is curved space with  
constant curvature $R_{(3)}=-\frac{24}{R^2}$.

To conclude this Section, let us note that deformations of the special relativity 
in some domain by means of the transformation $\Lambda_{def}=U^{-1}\Lambda U$ 
suggests the (singular) change of variables $X=U^{-1}x$. The variable $X$ has the standard 
transformation law under $\Lambda_{def}$: $X^{'}=\Lambda X$. It is true for the 
Fock-Lorentz realization [14] and for the recent DSR proposal [5] (see 
discussion in [14, 17]). Moreover, 
different DSR proposals in the momentum space can be considered either as different 
definitions of the 
conserved momentum $p^\mu$ in terms of the de Sitter momentum space 
variables $\eta^A$ [11], or as different definitions of $p^\mu$ in terms of the 
special 
relativity velocities $v^\mu=\frac{dx^\mu}{d\tau}$ [13]). Thus the known DSR 
proposals state, in fact, that experimentally measurable coordinates can be different 
from the ones specified as "measurable" by the special relativity theory.  
For the case under consideration, the transformation (\ref{3}) acts as  
ordinary Lorentz transformation on the variables
\begin{eqnarray}\label{7}
X^\mu\equiv\frac{x^\mu-\lambda x^2}{1+2\lambda x^0-\lambda^2x^2}. 
\end{eqnarray}
Geodesic motion of a particle in the space (\ref{6}) looks as a free motion in the 
coordinates (\ref{7}): $\ddot X=0$, see the next Section. So, Eq.(\ref{7}) 
represents coordinates of a locally inertial frame.

\section{Particle dynamics and kinematics on the DSR space.}

The invariant interval (\ref{6}) suggests the following  
action\footnote{In terms of the variables (\ref{7}) the Lagrangian acquires 
the form
$L=\frac{1}{2e}((\dot X(x))^2-em^2)$.} for a particle motion
\begin{eqnarray}\label{8}
S=\frac{1}{2}\int d\tau\left[\frac{\eta_{\mu\nu}\dot x^\mu\dot x^\nu}
{e(1+2\lambda x^0-\lambda^2x^2)^2}-em^2\right]. 
\end{eqnarray}
It is invariant under the global symmetry (\ref{3}), under the "translations": 
$x^{'\mu}=(SC_\lambda)^{-1}e^{a^{.}\partial}(SC_\lambda)x^\mu$ with the 
parameters $a^\mu$, 
as well as under the reparametrizations $\tau\longrightarrow\tau^{'}(\tau), ~ 
x^{'\mu}(\tau^{'})=x^\mu(\tau), ~ 
e^{'}(\tau^{'})=\frac{\partial\tau}{\partial\tau^{'}}e(\tau)$.
To discuss kinematics corresponding to the theory, it is convenient to use 
the Hamiltonian formulation for the system. One finds the canonical momenta 
for the variables $x, ~ e$
\begin{eqnarray}\label{9}
p^\mu=\frac{\dot x^\mu}{e\tilde\Omega^2}, \qquad p_e=0, 
\end{eqnarray}
and the Hamiltonian
\begin{eqnarray}\label{10}
H=\frac{1}{2e}(\tilde\Omega^2p^2+m^2)+\sigma_ep_e. 
\end{eqnarray}
Here and below the expressions of the type $p^2$ mean contraction with 
respect to the Minkowski metric $\eta_{\mu\nu}$. 
Transformation law for $p^\mu$ follows from (\ref{3})
\begin{eqnarray}\label{101}
p^{'\mu}=((\Lambda p)^\mu+
2\lambda(\delta^\mu{}_{0}-\Lambda^\mu{}_{0})(x p))G- \qquad \qquad \cr
2\lambda(p^0-\Lambda^0{}_{\nu}p^\nu-2\lambda(1-\Lambda^0{}_{0})(x p))
((\Lambda x)^\mu+\lambda(\delta^\mu{}_{0}-\Lambda^\mu{}_{0})x^2).
\end{eqnarray}
On the next step of the Dirac procedure, 
from the condition of preservation in time of the primary 
constraint: $\dot p_e=0$, one 
finds the secondary constraint, the latter represents deformed dispersion 
relation for the canonical momenta 
\begin{eqnarray}\label{11}
p^2=-\frac{m^2}{(1+2\lambda x^0-\lambda^2x^2)^2}.
\end{eqnarray}
There are no of tertiary constrains in the problem. Then dynamics for the 
variables (x, p) is governed by the equations 
\begin{eqnarray}\label{12}
\dot x^\mu=e\tilde\Omega^2p^\mu, \quad 
\dot p^\mu=-\frac{2em^2}{\tilde\Omega}(\delta^\mu{}_{0}\lambda+\lambda^2x^\mu), \quad 
p^2=-\frac{m^2}{\tilde\Omega^2}.
\end{eqnarray}
The equations acquire the most simple form in the gauge $e=\tilde\Omega^{-2}$ for the 
primary constraint $p_e=0$ (the gauge coincides with the standard one in the limit 
$\lambda\longrightarrow 0$)
\begin{eqnarray}\label{13}
\dot x^\mu=p^\mu, \quad 
\dot p^\mu=-\frac{2m^2}{\tilde\Omega^3}(\delta^\mu{}_{0}\lambda+\lambda^2x^\mu), \quad 
p^2=-\frac{m^2}{\tilde\Omega^2}.
\end{eqnarray}
The canonical momentum has now the standard expression in terms of 
velocity $p^\mu=\dot x^\mu$. As a consequence, it's transformation law coincides 
with the one for $\dot x^\mu$, see Eq.(\ref{4}).
The system (\ref{13}) implies the following Lagrangian equations for $x^\mu$
\begin{eqnarray}\label{14}
\ddot x^\mu+\frac{2m^2}{\tilde\Omega^3}(\delta^\mu{}_{0}\lambda+\lambda^2x^\mu)=0. 
\end{eqnarray}
The deformed gauge is convenient to study dynamics in a particular reference frame, 
but implies complicated law for transformation to other frames. Actually, 
to preserve the gauge, Eq.(\ref{3}) must be accompanied by reparametrization of the 
evolution parameter $\tau^{'}(\tau)$, where $\tau^{'}$ represents a solution 
of the equation 
$\frac{\partial\tau^{'}}{\partial\tau}=G^{-2}(\Lambda)$. In contrast, the gauge 
$e=1$ retains the initial transformation law (\ref{3}), and seem to 
be reasonable to discuss kinematics of the theory. 

Kinematical rules must be formulated for conserved energy and momentum. 
One notes that the 
canonical momentum (\ref{9}) is not a conserved quantity, in accordance 
with Eq. (\ref{12}). The discussion in the end of the Section 2 prompts that the 
conserved momentum may be $P^\mu=e^{-1}\dot X^\mu(x)$. It's expression in terms of the 
canonical momentum (in any gauge) is given by 
\begin{eqnarray}\label{15}
P^\mu=(p^\mu-2\delta^\mu{}_{0}\lambda (x p))\tilde\Omega-
2(x^\mu-\delta^\mu{}_{0}\lambda x^2)(\lambda p^0-\lambda^2(x p)). 
\end{eqnarray}
By direct computation one finds that $P^\mu$ is actually conserved 
on-shell (\ref{12})
and obeys the ordinary energy-momentum relation as a consequence of Eq.(\ref{11}). The 
deformed transformations (\ref{3}), (\ref{101}) induce the standard realization of 
the Lorentz group on $P^\mu$-space. As a consequence, composition 
rule for the momenta is 
the standard one, there is no problem of total momentum in the theory. 
So, the present version of the DSR theory leads to the standard kinematical rules 
on the space (\ref{15})   
\begin{eqnarray}\label{16}
\partial_\tau P^\mu=0, \qquad \eta_{\mu\nu}P^\mu P^\nu=-m^2\qquad 
\Lambda_\lambda: P^\mu\longrightarrow \Lambda^\mu{}_{\nu}P^\nu, \cr
P^\mu_{\sum}=\sum P^\mu_{(i)}. \qquad \qquad \qquad \qquad 
\end{eqnarray}
The energy and momentum have nonstandard relation (\ref{15}), 
(\ref{9}) with measurable quantities (velocities and coordinates). It suggests  
that kinematical predictions of the theory differ from that of the 
special relativity theory.

The difference among the canonical momentum and the conserved one implies an 
interesting situation in canonically quantized version of the theory. 
While the conjugated variables $(x, p)$ have the standard brackets, commutators 
of the coordinates $x^\mu$ with the energy and momentum $P^\mu$ are deformed,  
as it can be seen from Eq.(\ref{15}). Thus the phase space $(x, P)$ is endowed 
with the noncommutative geometry (with the commutators $[x, P]$ 
and $[P, P]$ being deformed). In 
particular, the energy-momentum subspace turns out to be noncommutative.
The modified bracket $[x, P]$ suggests that the Planck's constant has slight 
dependence on $x$
(similar bracket structure, with energy dependent Planck's constant, arises in 
the MS model [3]). 

\section{Conclusion}

In this work we have proposed version of the doubly special relativity theory in 
position space based on deformation of ordinary Lorentz transformations 
due to the special conformal transformation. There is unique deformation (\ref{3}) 
which does not modify the space rotations, namely, the deformation with  
the special conformal parameter being $b^\mu=(\lambda, 0, 0, 0)$. The invariant 
interval (\ref{6}) corresponds to the flat four-dimensional space-time (on a 
domain where the metric is non degenerated). By construction,  
maximum signal velocity is observer independent scale of the theory. 
The formulation 
admits one more independent scale $R\equiv -\frac{1}{\lambda}$, the latter 
is identified with 
radius of three-dimensional hypersection of (\ref{6}) at $x^0=0$. 

Geodesic motion of a particle on the space (\ref{6}) has been discussed in 
some details. The conjugated 
momentum $p^\mu$ (\ref{9}) for the coordinate $x^\mu$ has complicated 
transformation law (\ref{101}), and obeys the 
deformed energy-momentum relation (\ref{11}). The conserved energy-momentum $P^\mu$ 
(\ref{15}) turns out to be different from the canonical one. The transformations 
(\ref{3}), (\ref{4}) for $(x, p)$ induce the standard Lorentz transformations on 
the space of conserved momentum. It means, in particular, that composite rule 
for the energy-momentum is ordinary sum. There is no problem of total momentum 
in the theory. The conserved momentum, in contrast to the canonical one, obeys the 
standard energy-momentum relation. Kinematical rules of the theory are 
summarized in Eq.(\ref{16}). One expects that kinematical predictions of the 
theory differ from that of the special relativity due to 
nonstandard dependence of energy and momentum on measurable quantities, 
see Eqs.(\ref{15}), (\ref{9}).

\section{Acknowledgments}
Author would like to thank the Brazilian foundations CNPq and FAPEMIG 
for financial support.

\end{document}